%Paper: hep-th/9208050
%From: meurice@hepmips.physics.uiowa.edu (Yannick Meurice)
%Date: Wed, 19 Aug 92 21:52:31 -0500

%This works with the ias version of phyzzx.tex currently available via 'get'.
\input phyzzx.tex
%\voffset=3pc
%\hoffset=0.45in
%%%%%%%%%%%%%%%%%%%%%%%%%%%%%%%%%%%%%%%%%%%%%%%%%%%%%%%%%%%%%%%%%%%%%%%%%%%%%%%

%%REFERENCES%%%%%%%%%%%%%%%%%%%%%%%%%%%%%%%%%%%%%%%%%%%%%%%%%%%%%%%%%%%%%%%%%%
\def\PL  #1 #2 #3 {{\sl Phys.~Lett.}~{\bf#1} (#3) #2 }
\def\NP  #1 #2 #3 {{\sl Nucl.~Phys.}~{\bf#1} (#3) #2 }
\def\PR  #1 #2 #3 {{\sl Phys.~Rev.}~{\bf#1} (#3) #2 }
\def\PRD #1 #2 #3 {{\sl Phys.~Rev.~D} {\bf#1} (#3) #2 }
\def\PRB #1 #2 #3 {{\sl Phys.~Rev.~B} {\bf#1} (#3) #2 }
\def\PP  #1 #2 #3 {{\sl Phys.~Rep.}~{\bf#1} (#3) #2 }
\def\MPL #1 #2 #3 {{\sl Mod.~Phys.~Lett.}~{\bf#1} (#3) #2 }
\def\CMP #1 #2 #3 {{\sl Comm.~Math.~Phys.}~{\bf#1} (#3) #2 }
\def\PRL #1 #2 #3 {{\sl Phys.~Rev.~Lett.}~{\bf#1} (#3) #2 }
\def\TMP  #1 #2 #3 {{\sl Theor.~Math.~Phys.}~{\bf#1} (#3) #2 }
\def\JMP  #1 #2 #3 {{\sl Jour.~Math.~Phys.}~{\bf#1} (#3) #2 }
\def\IJ  #1 #2 #3 {{\sl Int.~Jou.~Mod.~Phys.}~{\bf#1} (#3) #2 }
%%%%%%%%%%%%%%%%%%%%%%%%%%%%%%%%%%%%%%%%%%%%%%%%%%%%%%%%%%%%%%%%%%%%%%%%
\REF\migdal{Detailed lists of references can be found e.g. in A. Migdal,
\PP 102 199 1983 or in A. Neveu, Les Houches 1982, J. Zuber and R. Stora,
Editors.}
\REF\polyakov{A. ~Polyakov, \PL 82B 247  1979  and  \PL 103B 211 1981 . }
\REF\kramers{H. Kramers and G. Wannier, \PR 60 252 1941 ; F.
Wegner, \JMP \hfil \break 12 2259 1971  ;
L. Kadanoff and H. Ceva, \PRB 3 3918 1971 ;
D. Merlini
and C. Gruber \JMP 13 1822 1972 ; R. Balian, J.M. Drouffe and C. Itzykson
\PRD 11 2098; A. Polyakov, \NP 120 429 1977 ; G. 't Hooft, \NP 138 1 1978 ;
E. Fradkin and L. Susskind \PRD 17 2637 1978 . }
\REF\dotpol{V. Dotsenko and A. Polyakov, in {\it Advanced Studies in Pure
Mathematics }, {\bf 16}, 1988.}
\REF\dotsenko{V. Dotsenko, \NP \hfil \break B285 45 1987 ; A. Kavalov
and A. Sedrakyan,
\NP 285 264 1987 ; P. Orland \PRL 59 2393 1987 ; J. Distler, preprint PUPT
1324. }
\REF\geneva{Y. Meurice,
Proceedings of the International Lepton-Photon Symposium 1991, $p.$114. }
\REF\wilson{K.~Wilson, \PRB 4 3174 1971 ; {\it ibid.} 3184. see also
K. Gawedzki
and A. Kupiainen, Les Houches 1985, K. Osterwalder and R. Stora, Editors.}
\REF\sinai{P. ~Bleher and Y. ~Sinai, \CMP 45 247 1975 ; P.~Collet and
J. P. ~Eckmann, \CMP 55 67 1977 and references therein. }
\REF\dyson{F. Dyson, \CMP 12 91 1969 .}
\REF\preprints{Y. Meurice, Univ. of Iowa Preprint. }
\REF\zabrodin{A. Zabrodin, \CMP 123 463 1989 . }
\REF\meurice{Y. Meurice, \PL 265B 377 1991 ; J.L. Lucio and Y. Meurice,
\MPL \hfil \break A6 1199 1991 . }
\REF\samuel{S. Samuel, \JMP 21 2806 1980 ; E. Fradkin, M. Srednicki and
L. Susskind, \PRD D21 2885 1980 ;
C. Itzykson, \NP 210 477 1982 . }

\REF\freund{P. Freund and M. Olson, \PL 199B 186 1987 ; P. Freund and
E. Witten \PL B199 191 1987  . }
%
%%%%%%%%%%%%%%%%%%%%%%%%%%%%%%%%%%%%%%%%%%%%%%%%%%%%%%%%%%%%%%%%%%%%%%%%%%
\Pubnum={UIOWA-91-27\cr
hep-th/9208050}
\titlepage
\title{ Remarks Concerning Polyakov's Conjecture for
the 3D Ising Model and  the Hierarchical Approximation}
\author{Yannick Meurice}
\address{Department of Physics and Astronomy, University of Iowa,
Iowa City, Iowa 52242, USA}
\vfil
\abstract
We consider the possibility of using the hierarchical approximation
to understand the continuum limit of a reformulation of the
$3D$ Ising model initiated by Polyakov.
We introduce several new formulations of the hierarchical model
using dual or fermionic variables.
We discuss several aspects of the renormalization group transformation in
terms of these new variables.
We mention a reformulation of the model closely related to
string models proposed by Zabrodin.

\endpage

\chapter{Introduction}

There has been a long-standing effort to describe the large distance
behavior of gauge theories in terms of a string theory.\refmark\migdal
This is obviously not an easy task and it would be very
interesting to have at hand a simple example where all
details can be carried out explicitly.
It seems worth trying with a 3$D$ $Z(2)$ gauge theory,
because this model is dual to the nearest neighbor $3D$
Ising model
for which we have a reasonable understanding of the critical behavior.
Polyakov\refmark\polyakov has shown that there exist linear
relations among
the averages of products of order and disorder variables
taken along contours. These reveal the existence of fermionic excitations
``held together" by gauge-invariance.
Subsequently, these contour averages have been reexpressed as
sums over ``equipped surfaces" by Polyakov and Dotsenko.\refmark\dotpol
These results strongly suggest that near the critical temperature,
the $3D$ Ising model can be described in terms
of some free fermionic string.

Despite significant efforts,\refmark\dotsenko we do not yet
have a clear understanding of
the continuum limit of the reformulation of Polyakov and Dotsenko.
For this reason, we have suggested\refmark\geneva
considering this problem in the context of the hierarchical approximation.
This approximation has been justified by Wilson\refmark\wilson
and used to calculate
the critical exponents of the 3$D$ Ising model with good precision.
The practical advantage of the hierarchical approximation is that the
renormalization group procedure reduces to the study of a simple integral
equation (``the approximate recursion formula"). The fixed points
of this recursion formula and the
relevant eigenvectors of the linearized transformation have been studied
in great
detail.\refmark\sinai
The critical
exponents have a very simple form in terms of the corresponding
eigenvalues.

The hierarchical approximation holds exactly in the case of the
hierarchical model\refmark\dyson (HM in the following) which will
be the main object of the present study. For definiteness, this model
is briefly reviewed in section 2. We then introduce
several new formulations of the HM.
The model dual to the HM
is described in section 3. Various fermionizations are discussed
in section 4. Due to the fact that the model has long range interactions,
specific methods were necessary to obtain these results.
These methods indeed apply to a larger class
of models and are reported with due mathematical care in a separate
article.\refmark\preprints
Finally, we mention in the conclusion that that the
model can be rewritten as a nearest neighbors model on a ``branch"
by introducing auxiliary (gaussian) fields.
This construction is related, but distinct on several points, to
string models proposed by Zabrodin.\refmark\zabrodin

The main question we address is how the
extraordinary simplification provided by the renormalization group method
(we only need the relevant directions)
takes place when, instead of the order
variables, we use the new variables discussed above.
In all the cases, the renormalization
group transformation appears to be more involved than
in the usual formulation. This can be understood from the fact that
in the HM, the number of links grows proportionally to the square
of the number of sites. However, preliminary results indicate that
when the symmetries of the model
are taken into account, the procedure can be reduced to a size
comparable to the usual one.
The hard work (fixed points, linearization) remains to be done.
A proper understanding of this question may shed a new light
on the continuum limit of the sum over equipped surfaces discussed
above.

\chapter{The Hierarchical Model}

For definiteness recall a few facts
concerning the HM.\refmark\dyson
More details and references can be found in the lecture notes
of Collet and Eckmann\refmark\sinai or Gawedzki and Kupiainen.\refmark\wilson
The hamiltonian of a hierarchical model with $q^n$ sites can be written as
$$H=-{1 \over 2} \sum\limits_{l=1}^{n}({c \over q^2})^l
\sum\limits_{i_n,...,i_{l+1}}\ (\ \sum\limits_{i_l,....,i_1} \sigma
_{(i_n,....,i_1)} )^2\ \eqno(1) $$
All the indices
$i_j$ run from 1 to $q$, an integer which controls the size of the boxes
used during the renormalization group transformation.
We assume that
$1\leq c<q$. It is suggestive to write
$c=q^{1-d_h}$ where $d_h$ is the Hausdorff dimension\refmark\meurice
of the random walk associated with $H$. If the spins are integrated with a
gaussian measure,
the model has the same critical exponents as
an ordinary gaussian model in $D$ dimensions, provided that $d_h={2 \over D}$.
A non-gaussian continuum
limit exists for $1> d_h>1/2$.

In the next two sections, we use the short notation
$i$ instead of $(i_n,....,i_1)$ and we define a function $v(i,j)$
which is equal to $l$ if $i_l$ is the first index (starting from the left)
differing in $i$ and $j$. Another way of saying it that
the smallest box containing $i$ and $j$ contains $q^{v(i,j)}$
sites. The partition function reads
$$ Z=\sum_{\{ \sigma _i = \pm 1 \} }
e ^{\sum\limits_{i<j} K_{ij}\sigma _i\sigma_j}\ . \eqno(2) $$
\def\cp{{c \over q^2}}with
$$K_{ij}=\beta (1-\cp)^{-1}((\cp )^{v(i,j)}-(\cp)^{n+1})\ . \eqno(3)$$
The notation $i<j$ in the sum means that $i$ is distinct from $j$ and that
the pairs of distinct elements which can
be obtained by interchange are only counted once.

\chapter{The Duality Transformation}

As we have seen, the Kramers-Wannier duality\refmark\kramers
plays an important role in Polyakov's approach. In the case of
the HM,
a dual spin $S_{(ijk)}$ is  associated with any triangle $ijk$.
These triangles can be used to build surfaces whose boundary is a contour
appearing in the high temperature expansion of the model. The fact that
tetrahedrons are closed surfaces provides the principle underlying the
gauge-invariance of the dual formulation. All we need to check is that
all the terms of the high temperature expansion appear with
the same ``gauge-multiplicity''. This is proven elsewhere.\refmark\preprints
In the following, we only state the main results.

If we define the dual couplings
$tanh K_{ij}=e^{-2D_{ij}}\ $, we can write the partition function of Eq.(2) as
$$Z=(\prod_{i<j} coshK_{ij})\ 2^{N-{N-1 \choose 3 }}
\sum_{\{ S_{ijk}= \pm 1\}}
e^{+ \sum\limits_{i<j} (D_{ij} (\prod\limits_{k \ne i,j}
S_{(ijk)} -1))} \  .\eqno(4)$$
This reformulation has a local invariance.
For any four distinct sites $i,j,k$ and $l$, the dual
interaction energy is invariant under the simultaneous changes
of sign of $ S_{(ijk)},S_{(ijl)},S_{(ikl)}$ and $S_{(jkl)}$.
A possible gauge-fixing condition is $S_{(ijk)}=0$ if none of $i,j$
and $k$ are equal to a given site. In general,
if we start with $q^n$ sites, the dual formulation has
$(1/2)(q^n-1)(q^n-2)$ physical degrees of freedom.
This number can be obtained directly
by subtracting the $q^n -1$ independent
constraints on the parity of the number of visits
at each site from the total number of links.

We insist on the fact that the symmetries of the model\refmark\dyson are
crucial
to deal with this proliferation.
In order to fix the ideas, let us just compute the
high temperature expansion of the partition function ``hierarchically''.
We perform successive integrations over
the links variables inside boxes of size
$q^l$, starting with $l=1$ and proceeding similarly for higher $l$.
At first sight it looks like we have to keep track of a prohibitively
large number of
quantities, namely $2^{q^l}$, since each site can be visited
an even or an odd number of times
and we need this information when we integrate over the links in larger boxes.
At the end of the calculation (say for $l=n$), we only retain
those contributions where the number
of visits at each site is even.
This indeed amounts to calculate ${\it all}$ the correlation functions for
$q^l$ sites for $l=1,2..,n-1$. Fortunately, when the
symmetries of the model are taken into account,
the number of  correlation functions to calculate is reduced logarithmically.
It seems thus, in principle possible to use the
renormalization group procedure
in the dual formulation. Nevertheless, for practical reasons, it seems
advantageous
to first represent the high temperature expansion as
an integral over Grassmann variables\refmark\samuel.

\chapter{Hierarchical Fermions}

\def\psij{\psi _i ^j}
\def\psj{\psi _j ^i}
\def\chij{\chi _i ^j}
\def\chj{\chi _j ^i}
\def\meas{2^N \int [d\psi d\chi]}
\def\dpdc{\prod\limits_{i<j}d\chij d\psij d\chj d\psj }

A compact formulation of the high temperature expansion of the HM can be
obtained\refmark\preprints \break
by associating to each link $ij$ the four
Grassmann variables $\psij,\psj,\chij$ and $\chj$.
The final result
for the partition function is
$$Z=(\prod_{i<j} coshK_{ij}) \meas  \ e^{\ \sum\limits_{i<j}
((thK_{ij}) \psij \psj-\chij \chj )} \ \prod\limits_{i=1}^N z(\sum
\limits_{j;j\neq i} \psij \chij)\eqno(5)$$
where
$[d\psi d\chi ] \ = \ \dpdc $ and $z(x)$ is equal to
$sinh(x)$ (resp. $cosh(x)$)
if $q$ is even (resp. odd). The average value of products of order or
disorder variables and the Schwinger-Dyson equations are easily obtainable
from Eq.(5).
Note that, unlike in the 3D Ising model,
these equations are highly non-linear.
Note also that the number of Grassmann variables introduced is not minimal.
We have doubled the number of fermions in order to avoid the
splitting of the partition
function into an uncontrollably large number of terms. The details concerning
the above results can be found
in Ref.[\preprints].

This representation seems well-suited for the renormalization group approach.
We can proceed in two steps. First, we integrate over
the variables corresponding
to the links inside the boxes of size $q$. Second, we make a linear
transformation among the
$4 q^2$ variables associated to the links joining any pair of boxes of
size $q$ in
such a way that the average over upper and lower indices for the four type of
variables are among the new variables. We then perform a gaussian integration
over the remaining ones. After an appropriate rescaling, the ``kinetic term''
keeps its original form. What is transformed is the function $z(x)$ appearing
in Eq.(5). We intend to study this transformation
with the  methods used for the recursion formula mentioned in the introduction.

Note also that we can construct a fermionic representation of the
low temperature
expansion. To each dual site $ijk$ we associate the three Grassmann variables
$\psi _{ij} ^k ,\psi _{ik} ^j, \psi _{jk} ^i $ and their $\chi$ counterpart.
The kinetic term contains a part of the form
$$\sum\limits_{i<j} th D_{ij} \prod\limits_{k;k\neq i,j}
\psi _{ij} ^k \ \eqno(6)$$
and the method based on gaussian integration is not applicable.

\chapter{Conclusions and Perspectives}

We have introduced several reformulations of the hierarchical model.
The renormalization
group method seems an appropriate tool to handle at least one of them.
We think that a study of the fixed points and the relevant directions
of the new transformation will provide a better understanding of the continuum
limit of the reformulation of the $3D$ Ising model proposed
by Polyakov and Dotsenko.
This issue may also be clarified by a more systematic understanding of the
hierarchical approximation. This could be done by using a complete
set of multiplicative
characters and taking the degree of ramification as the order of perturbation.

Note also that the HM can be reformulated as a
nearest-neighbor model on a ``branch''. The hamiltonian reads
$$H= \sum\limits_{l=1}^{n}
\sum\limits_{i_n,...,i_{l+1}}
\ (\ {a_l \over 2}(\sigma_{(i_n,....,i_{l+1})}^{(l)} )^2
-{c^{1/2} \over q} \sigma_{(i_n,....,i_{l+1})}^{(l)}\sum\limits_{i_l}
\sigma_{(i_n,....,i_l)}^{(l-1)})
\eqno(7)$$
with
$a_1=1$ and
$a_l=1+{c \over q}$ for $ l\geq 2$.
The $\sigma ^{(0)}_i$
are the spins variables of the HM as in section 2.
The new variables $\sigma ^{(l)}$, for $l \geq 2$ are integrated with
$(2\pi \beta)^{-1/2}\int\limits_{- \infty}^{ +\infty} d\sigma ^{(l)}_i$.
This construction has been inspired by Zabrodin's models\refmark\zabrodin
designed to reproduce the $p$-adic generalization\refmark\freund
of the tachyon amplitudes of string theory.
Our reformulation differs on several points (discreteness of
the boundary, the local
couplings are not necessarily the same) from these models, however,
it might be worth exploiting
their resemblances.
The ``vertex operator''
$V_k=\sum\limits_i Exp(k \sigma ^{(0)}_i)$
can be used to generate the average value of functions of the total spin.
Note however, that in the gaussian case,
the $SL_2$ invariance\refmark{\freund,\zabrodin}
applies only for $d_h$=1 i.e $D=2$. Otherwise, a ``mass term''
spoils the transformation
properties and we cannot use the Koba-Nielsen trick.

\ack
We thank F. Goodman, W. Klink, A. Kupiainen, Y. Nambu, P. Orland
and C. Zachos for valuable
discussions. This paper is dedicated to the memory of our friends and
colleagues Dwight Nicholson,
Bob Smith, Chris Goertz and Linhua Shan.

\refout
\bye
\end